\documentstyle[prl,twocolumn,aps,epsf]{revtex}
\def\img{\mbox{i}}
\def\ostr{O_{\mbox{str}}}
\def\JF{J_{\mbox{F}}}
\def\H{{\cal H}}
\def\v#1{\mbox{\boldmath$#1$}}
\def\simleq{\mbox{\raisebox{-1.0ex}{$\stackrel{<}{\sim}$}}}

\begin{document}
\draft
\title{Density Matrix Renormalization Group Study\\ of the Haldane Phase in
Random One-Dimensional Antiferromagnets}
\author{Kazuo Hida}
\address{Department~of~Physics,
 Faculty of Science,\\ Saitama University, Urawa, Saitama 338-8570, JAPAN}

\date{Received 24 February 1999}

\maketitle

\begin{abstract}
It is conjectured that the Haldane phase of the $S=1$ antiferromagnetic
Heisenberg chain and the $S=1/2$ ferromagnetic-antiferromagnetic alternating
Heisenberg chain is stable against any strength of randomness, because of
imposed breakdown of translational symmetry. This conjecture is confirmed
by the density matrix renormalization group
calculation of the string order parameter and the energy gap distribution.
\end{abstract}

\pacs{75.10.Jm, 75.40.Mg, 75.50.Ee, 75.50.Lk}

In the recent studies of quantum many body problem, the ground state properties
of the random quantum spin systems have been attracting a renewed
interest
\cite{dg1,bl1,df1,kh1,hy1,bsc1,cfg1,hy2,kh3,mon1,yn1,yn2,tkt1,ren1,ha1}.
 Among them, the effect of randomness on the spin gap state of
 quantum spin chains has been extensively studied
theoretically and experimentally
\cite{hy1,bsc1,cfg1,hy2,kh3,mon1,yn1,yn2,tkt1,ren1,ha1}.

The real space renormalization group (RSRG) method has been often used for the
study of random quantum spin chains. Using this method, it has been exactly
proved that the ground state of  the $S=1/2$ random antiferromagnetic
Heisenberg chain (RAHC) is the random singlet(RS) state\cite{dg1,bl1,df1}
irrespective of the strength of randomness. Hyman et al.\cite{hy1} have applied
this method to the $S=1/2$ dimerized RAHC and have shown that the dimerization
is relevant to the RS phase. They concluded that the ground
state of this model is the random dimer (RD) phase in which the string long
range order survives even in the presence of randomness\cite{hy1,hy2}. These
results are numerically confirmed using the density matrix renormalization
group (DMRG) method\cite{kh1,kh3}.

The effect of randomness on the Haldane
phase is also studied by Boechat and coworkers\cite{bsc1,cfg1} and Hyman and
Yang\cite{hy2} using the RSRG method for the original model and the low energy
effective model, respectively. These authors predicted the possibility of the
RS phase for strong enough randomness. This problem has been further studied by
Monthus and coworkers using the numerical analysis of the RSRG equation for the
square distribution of exchange coupling\cite{mon1}. They predicted that the
Haldane-RS phase transition takes place at a finite critical strength of
randomness. In the finite neighbourhood of the critical point, the Haldane
phase belongs to the Griffith phase with finite dynamical exponent $z > 1$.
Hereafter this phase is called the random Haldane (RH) phase. On the other
hand, Nishiyama\cite{yn1} has carried out the exact diagonalization study of
the $S=1$ RAHC. He observed that the Haldane phase is quite robust against
randomness and the string order remains finite unless the bond strength is
distributed down to zero. He also carried out the quantum Monte Carlo
simulation\cite{yn2} and found no random singlet phase even for strong
randomness. On the contrary, the quantum Monte Carlo simulation by Todo {\it et
al.}\cite{tkt1} suggested the presence of the RS phase for strong enough
randomness.

In the absence of randomness, the present author has given a physical picture
of the Haldane phase as the limiting case of the $S=1/2$ Heisenberg chain with
bond alternation in which the exchange coupling takes two different values $J$
and $\JF$ alternatingly\cite{kreg1}. In the extreme case of $\JF \rightarrow
-\infty$, this system tends to the $S=1$ antiferromagnetic Heisenberg chain.
The string order remains long ranged over the whole range $-\infty < \JF < J$
and only vanishes at $J=\JF$. The perfect string order is realized for $\JF=0$.
As discussed by Hyman et al \cite{hy1}, this is the direct consequence of the
imposed breakdown of translational symmetry. Because the randomness cannot
recover the translational symmetry, the string order is expected to remain
finite over the whole range  $-\infty < \JF < J$ for any strength of
randomness. Therefore we may safely conjecture that the Haldane phase of the
$S=1$ RAHC should also remain stable for any strength of
randomness.

In the following, we confirm this conjecture using the DMRG
method\cite{kh1,wh1} which allows the calculation of the ground state and low
energy properties of large systems with high accuracy. We use the algorithm
introduced in
 ref. \cite{kh1}. This method has been successfully applied to the spin-1/2
RAHC and weakly dimerized spin-1/2 RAHC in which the system is gapless or has very small
gap in the absence of randomness. Namely, in these systems the characteristic
energy scale of the regular system is much smaller than the strength of
randomness even for weak randomness. Compared to these examples, present model
is less dangerous because the regular system has a finite gap and the
characteristic energy scale of the regular system is comparable to the strength
of randomness even in the worst case.
We investigate not only
the $S=1$ RAHC but also the $S=1/2$ random ferromagnetic-antiferromagnetic
alternating Heisenberg chain (RFAHC) which interpolates the $S=1/2$ dimerized
RAHC and the $S=1$ RAHC.

The Hamiltionan of the $S=1/2$ RFAHC is given by,
\begin{equation}
\label{eq:ham}
\H = \sum_{i=1}^{N} 2J_i\v{S}_{2i-1}\v{S}_{2i}+
2J_{\mbox{F}}\v{S}_{2i}\v{S}_{2i+1},\ \ \mid \v{S}_{i}\mid = 1/2,
\end{equation}
where $J_{\mbox{F}}=\mbox{const.}$ and  $J_i$'s are distributed randomly with
probability distribution,

\begin{equation}
P(J_i)=
\left\{\matrix{1/W & \mbox{for}\ 1-W/2 < J_i <1+W/2 \cr
0 & \mbox{otherwise.}}
\right.
\end{equation}
The width $W$ of the distribution represents the strength of randomness.
The maximum randomness is defined by $W=2$, because the ferromagnetic bonds
 appear among $J_i$'s for $W > 2$. It should be noted that the appearance
 of the random ferromagnetic bonds can drive the system to the completely
 different fixed point called large spin phase\cite{wes1}. Although the
 crossover between the random Haldane phase and the large spin phase is an
 interesting issue, we leave this problem outside the scope of the present
study.

The ground state of the regular counterpart of this model ($W=0$) is the
Haldane phase with long range  string order defined by
$\ostr = \lim_{l, N \rightarrow \infty} \ostr(l;N)$\cite{kreg1}.
Here $\ostr (l;N)$ is the string correlation function in the chain of length
$N$ defined only for odd $l$ as,
\begin{equation}
\label{eq:ost2}
\ostr (l;N) = -4<\exp \left\{\img\pi \sum_{k=2i+1}^{2i+l+1} S_{k}^z \right\} >.
\end{equation}
where $<...>$ denotes the ground state average. In the presence of randomness,
the string order is defined as the sample average of $\ostr$.
In the limit $J_{\mbox{F}} \rightarrow -\infty$, the string
order parameter (\ref{eq:ost2}) reduces to the one for the $S=1$
chain\cite{dnr,tasaki}.  

First, we calculate $\ostr$ for the $S=1/2$  RFAHC and $S=1$
RAHC using the DMRG method.
The calculation is performed with open boundary condition. For the $S=1/2$
RFAHC,
 the bonds at the both ends of the chain are chosen to be antiferromagnetic to
avoid
the quasi-degeneracy of the ground state. The two
boundary spins are not counted in the number of spins $2N$ to keep the
consistency with the $S=1$ chain (see below). The average is taken over 200
samples with $N \leq 29$ (58 spins). The string order for the finite system
$\ostr(N)$ is estimated from $\ostr(l; N )$ averaged over 6 values of $l$
around $l=N/2$. The maximum number $m$ of the states kept in each step is 100.
Similar calculation is also performed for the $S=1$ RAHC. In this case, the
additional spins with $S=1/2$ are added at the both ends of the chain to remove
the quasi-degeneracy as proposed by White and Huse for the regular
chain\cite{wh3}. The average is taken over 400 samples with $N \leq 42$ where
$N$ is the number of $S=1$ spins. In this case, we take $m=80$ and  $\ostr(N)$
is estimated from $\ostr(l; N )$ averaged over 12 values of $l$ around $l=N/2$.
We have confirmed that these values of $m$ are large enough from the
$m$-dependence of the obtained values of $\ostr(N)$. 
\begin{figure}
\epsfxsize=55mm 
\centerline{\epsfbox{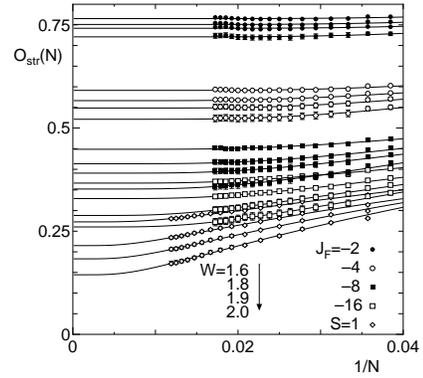}}
\caption{The size dependence of the string order parameter $\ostr (N)$ plotted
against $1/N$. The solid lines are fit by the formula (6). The points with
different values of $W$ are depicted by the same symbols for $W=1.6$, 1.8, 1.9
and 2.0 from top to down.}
\label{fig1}
\end{figure}
\begin{figure}
\epsfxsize=55mm 
\centerline{\epsfbox{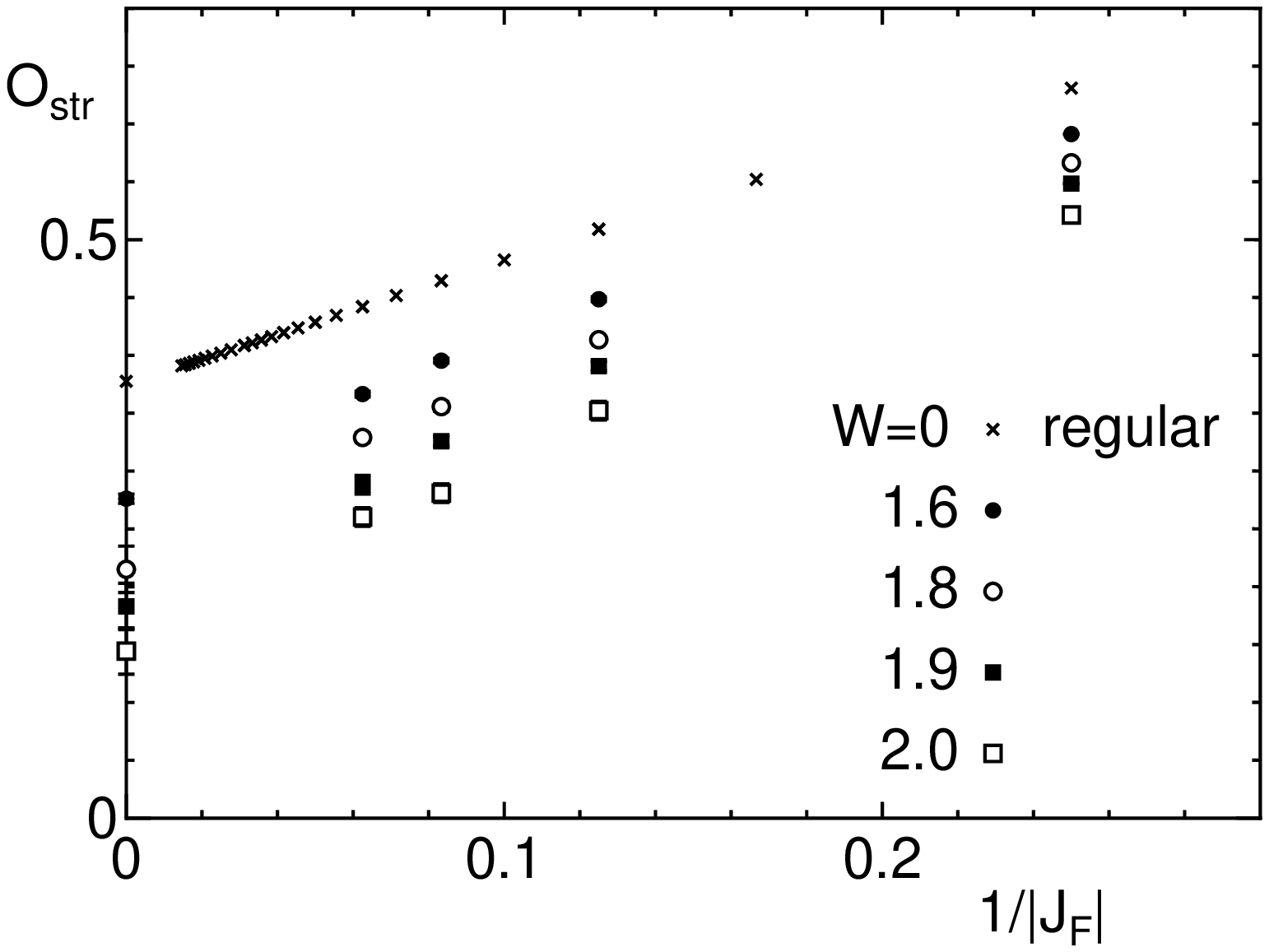}}
\caption{The $J_{\mbox{F}}$-dependence of the string order parameter $\ostr $
for $W=1.6$, 1.8, 1.9  and 2.0 plotted against $1/\mid J_{\mbox{F}} \mid$. }
\label{fig2}
\end{figure}

Figure \ref{fig1} shows the size dependence of the string order. Typical sizes of the error bars estimated 
from the statistical flucuation among samples 
are less than the size
 of the symbols unless they are explicitly
shown in the figures. The
extrapolation is made under the assumption
\begin{equation}
\ostr(N) \simeq \ostr +CN^{-2\eta} \exp(-N/\xi).
\label{extra}
\end{equation}
where $C$ and $\xi$ are the constants to be determined by fitting. The exponent
$2\eta$ characterize the size dependence of the string order parameter at the
RH-RS critical point where $\xi$ should diverge as $\ostr(N) \sim N^{-2\eta}$.
This value is estimated as follows: According to Monthus et al.\cite{mon1}, at
the critical point, $\ostr(N)$ behaves as $\Gamma^{-2(3-\phi)}$ with $\phi =
\sqrt{5}$ while the logarithmic energy scale $\Gamma$ varies with the system
size as $\Gamma \sim N^{1/3}$. Therefore $\ostr(N)$ should scale as
$N^{-2(3-\phi)/3} \sim N^{-0.5092}$ at the critical point resulting in $2\eta =
0.5092$.  It should be noted that the low energy effective model of Hyman and
Yang\cite{hy2} also applies for the $S=1/2$ RFAHC with finite $J_{\mbox{F}} <0
$ by construction. 
The extrapolated values of $\ostr$ are plotted against
$1/\mid \JF \mid$  in Fig. \ref{fig2}. 
The string order is perfect at
$J_{\mbox{F}}=0$, where
the ground state is a simple assembly of local singlets\cite{kreg1} and should
decrease
with the increase of $\mid J_{\mbox{F}}\mid$. This behavior is clearly seen
in Fig \ref{fig2}. In this extrapolation scheme, the string
long range order remains finite even at $W =2$ for both $S=1/2$ RFAHC and $S=1$ RAHC.
\begin{figure}
\epsfxsize=55mm 
\centerline{\epsfbox{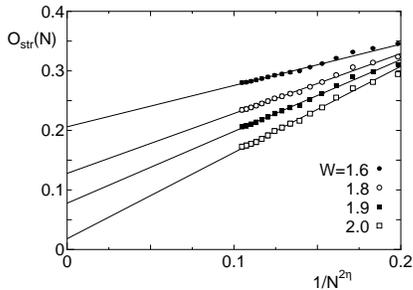}}
\caption{The size dependence of the string order parameter $\ostr (N)$ for the
$S=1$ RAHC plotted against $N^{-0.5092}$.}
\label{fig3}
\end{figure}
\begin{figure}
\epsfxsize=55mm 
\centerline{\epsfbox{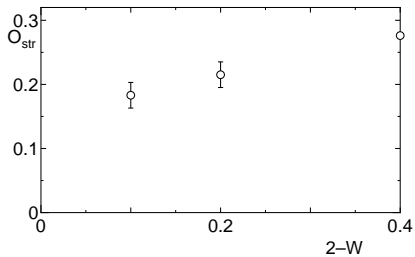}}
\caption{The $W$-dependence of the string order parameter $\ostr$ for the $S=1$
RAHC.}
\label{fig4}
\end{figure}

To check our extrapolation scheme, we also made the extrapolation assuming the
size dependence $N^{-2\eta}$ expected at the RH-RS critical point in  Fig.
\ref{fig3} for $S=1$ RAHC. In the RH phase, the extrapolated values thus
obtained can be understood as the lower bound. For $W \geq 1.9$, it is clear
that the extrapolated values remain definitely positive. Therefore the
extrapolation using Eq. (\ref{extra}) is appropriate in this region rather than
the power law extrapolation. For $W=2$, the extrapolated
value is  very small but still positive ($\simeq 0.018$).

Even if we do not rely on the values  for $W=2$ extrapolated using Eq.
(\ref{extra}), we can convince ourselves the stability of the RH phase at $W=2$
by the following argument.  In Fig. \ref{fig4}, we plot $\ostr$ extrapolated
using  Eq. (\ref{extra}) for $S=1$ RAHC against $2-W$ for $W \leq
1.9$. If we assume the critical behavior $\ostr \sim (W_c-W)^{1.173}$
predicted by Monthus et al.\cite{mon1}, it is highly unlikely that the string
order disappears at finite values of  $W_c$ less than 2.

Especially, this excludes the possibility $W_c
\simeq 1.485$ predicted by Monthus et al.\cite{mon1}. In general, it is not surprising that
the RSRG method gives incorrect value for the critical point even if it gives
correct values for the critical exponents, because the RSRG transformation is
not exact at the initial stage of renormalization. Furthermore, Monthus et al.\cite{mon1} have neglected the effective ferromagnetic coupling between the next nearest neighbour interaction which appear after decimation of two $S=1$ spins. (See the discussion following eq. (2.20) of ref. \cite{mon1}.) The neglect of this term is equivalent to the introduction of the antiferromagnetic next nearest neighbour interaction as a counter term in the bare interaction. In terms of the RFAHC, such interaction makes the distinction between the even and odd bonds meaningless and can recover the translational symmetry leading to the destruction of the string order erroneously.
\begin{figure}
\epsfxsize=55mm 
\centerline{\epsfbox{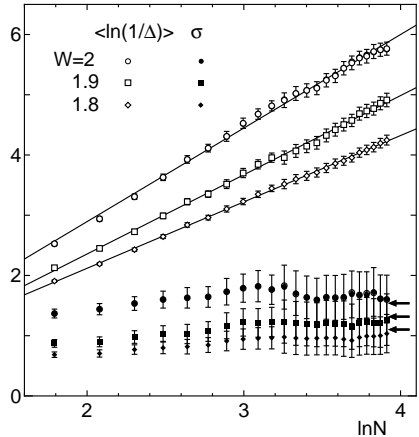}}
\caption{The $N$-dependence of $<\ln(1/\Delta)>$ (open symbols)  and $\sigma$
(filled symbols) of the $S=1$ RAHC plotted against $\ln N$ for $W=1.8$, $1.9$
and 2.0. The horizontal arrows indicate the values of $1/P_0$ estimated from
$d<\ln(1/\Delta)>/d \ln N $ for $W=2, 1.9$ and 1.8 from top to
bottom.}
\label{fig5}
\end{figure}
To further confirm the stability of the RH phase at the most dangerous point
$\JF \rightarrow -\infty$ and $W \simleq 2$, we calculate the energy gap
distribution for the $S=1$ RAHC.
 In the RH phase, the fixed point distribution
of the energy gap $\Delta$ is given by
$P(x) = P_0 \exp(-P_0x)$
where $x \equiv \ln (\Omega/\Delta)$, $\Omega$ is the energy cut-off and $P_0$
is the nonuniversal constant\cite{hy2}. For the finite size systems, $\Omega $ scales as
$N^{-1/P_0}$\cite{hy2}. Therefore the  dynamical exponent $z$ is given by $z
=1/P_0$. Furthermore, this distribution implies
\begin{equation}
<\ln(1/\Delta)> = P_0^{-1}\ln N + \mbox{const.} 
\end{equation}
\begin{equation}
\sigma \equiv \sqrt{<(\ln \Delta-<\ln \Delta >)^2>} = P_0^{-1}=z
\end{equation}
for $N >> 1$. In Fig. \ref{fig5}, we plot $<\ln (1/\Delta) >$ and $\sigma$
against $\ln N$. The error bars are estimated from the statistical flucutation among samples. 
The average is taken over more than 100 samples with $N \leq
50$. For the most random case $W=2$, the average is taken over 219 samples. In
this case, we have taken $m=100$ in most cases. For the confirmation of the
accuracy, however, we recalculated with $m=150$ for the samples with very small
gap (less than $10^{-3}$) but the difference was negligible. Actually, the latter data
are also plotted in Fig. \ref{fig5} for $W=2$. But they are almost covered by the data with $m=100$ 
and are invisible in Fig. \ref{fig5}. Therefore we
may safely neglect the $m$-dependence for less dangerous case $W < 2$.

\begin{figure}
\epsfxsize=55mm 
\centerline{\epsfbox{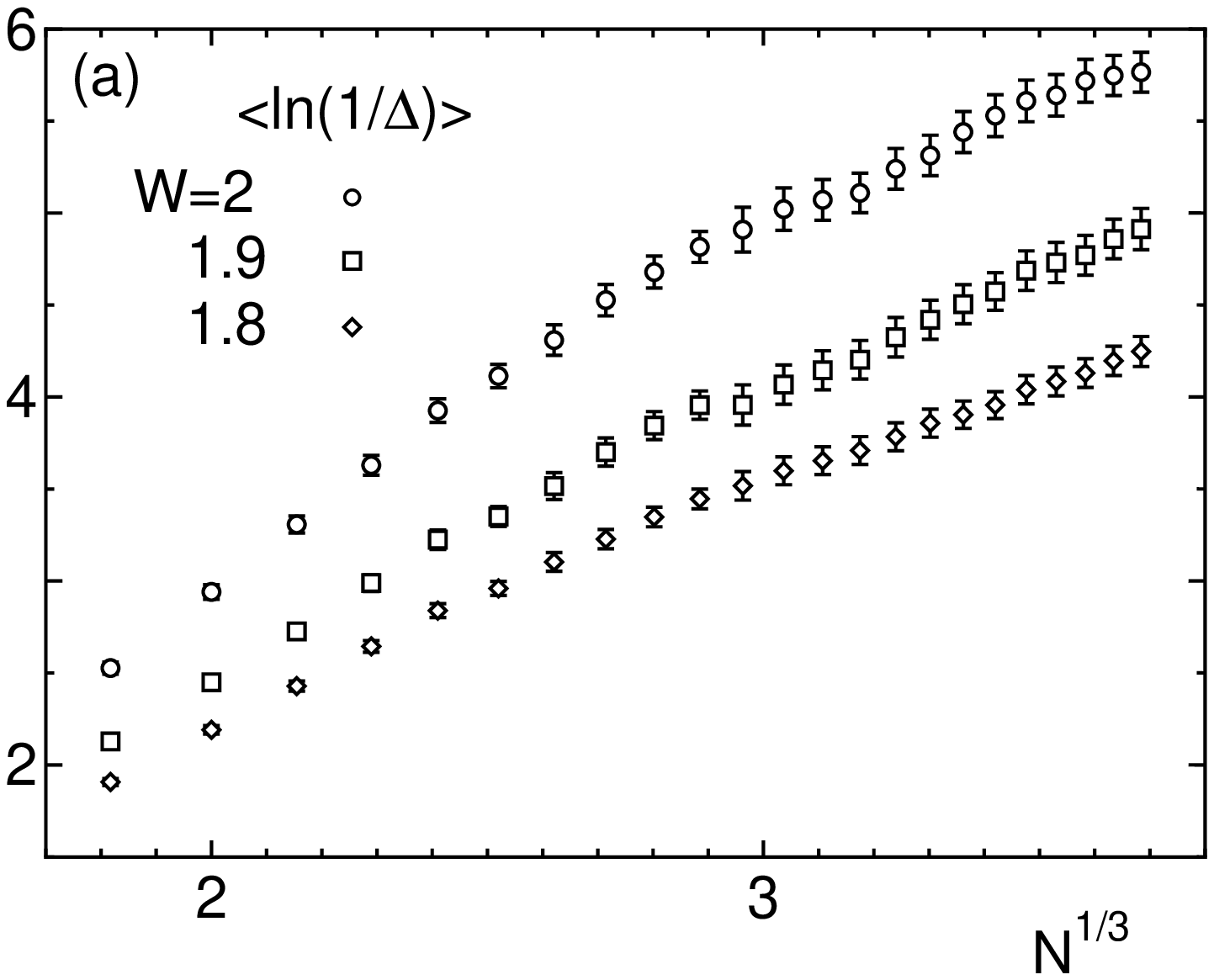}}
\epsfxsize=55mm 
\centerline{\epsfbox{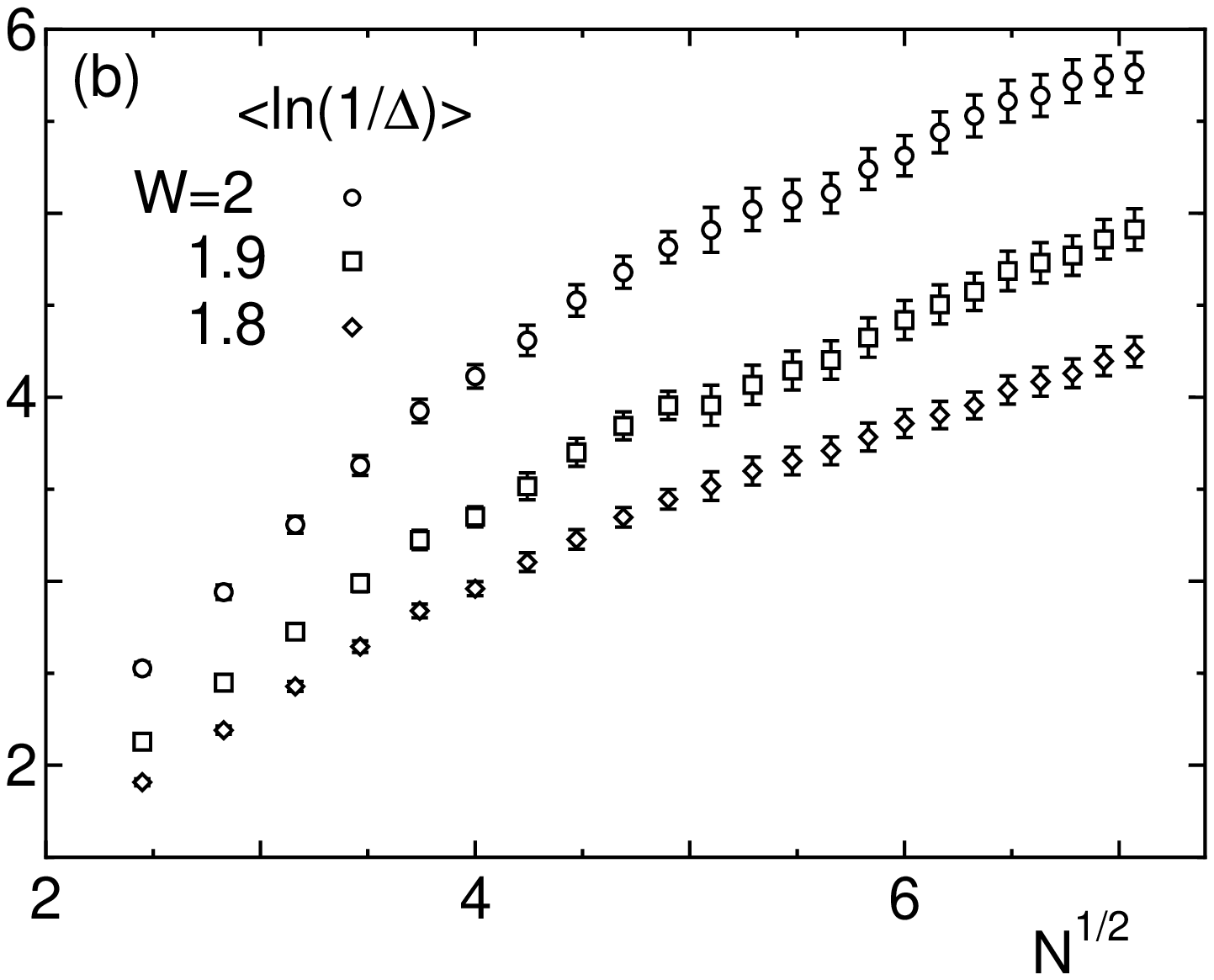}}
\caption{The $N$-dependence of $<\ln(1/\Delta)>$ of the $S=1$ RAHC plotted (a)
$N^{1/3}$ and  (b) $N^{1/2}$ for $W=1.8$, $1.9$ and 2.0.}
\label{fig6}
\end{figure}

It is evident that  $<\ln (1/\Delta) >$ behaves linearly with $\ln N$ and
$\sigma$ tends to a constant value as $N \rightarrow \infty$. We can estimate
the values of $z (=1/P_0)$ from the gradient  $d<\ln(1/\Delta)>/d\ln N$ for
large $N$. These values are indicated by the arrows in  Fig. \ref{fig5} and
they are consistent with those estimated from $\sigma$ for large $N$ within the
error bars. On the other hand,  Figs. \ref{fig6}(a) and (b) show the plot of
$<\ln(1/\Delta)>$  against $N^{1/3}$ and $N^{1/2}$ which are the expected size
dependence at the RH-RS critical point and within the RS phase,
respectively\cite{mon1}. Both plots are less linear compared to Fig.
\ref{fig5}. These results confirm that the ground state of the $S=1$ RAHC
remains in the RH phase down to $W=2$.

It should be also noted that the finite size effect becomes serious only if one
hopes to conclude the {\it presence} of the RS phase.
 Even in the RH phase, the string order
or the gap distribution might behave RS-like if the system size is not enough.
Actually, the authors
of ref. \cite{mon1} needed extremely large number of spins to conclude that
their calculation leads to
the RS phase. But the RS phase can never behave RH-like by the finite size effect because the RS phase has divergent correlation length. Therefore, it is relatively easy to {\it exclude} the possibility
of RS phase if the
deviation from the RS-like behavior is already observed for relatively small
systems, which is the
case of the present calculation. 

In  summary, it is conjectured that the Haldane phase of the $S=1$ RAHC and the $S=1/2$ 
RFAHC is stable against any strength of randomness, because of the 
imposed breakdown of translational symmetry. This conjecture is confirmed by the DMRG calculation of the string order and the energy gap distribution. 


The numerical calculations have been performed using the FACOM VPP500 at the
Supercomputer Center, Institute for Solid State Physics, University of Tokyo.
The author thanks H. Takayama and S. Todo for useful discussion and comments. This work is partially supported by the Grant-in-Aid for Scientific Research from the Ministry of Education, Science, Sports and Culture.

\end{document}